\documentclass[aps,twocolumn,showpacs,pra,superscriptaddress]{revtex4-1}
\pdfoutput=1

\usepackage{graphicx}
\usepackage{epstopdf}
\usepackage{amsmath,amssymb}
\usepackage{upgreek}
\usepackage{fancyhdr}
\usepackage[T1]{fontenc}
\usepackage[utf8]{inputenc}
\usepackage[colorlinks, linkcolor=blue, citecolor=blue, urlcolor=blue,breaklinks=true]{hyperref}
\usepackage{natbib}
\usepackage{soul}
\usepackage{subfigure}

\graphicspath{{figures/}}

\newcommand{\braket}[2]{\left\langle #1\left |\vphantom{{#1}{#2}} 
\right.#2\right\rangle}
\newcommand{\rr}{{\boldsymbol r}}
\newcommand{\sub}[1]{{\textrm{#1}}}

\providecommand{\braket}[2]{\left\langle#1\left|\vphantom{{#1}{#2}}\right.#2\right\rangle}

\begin{document}

\title{Magnetic noise spectrum measurement by an atom laser in gravity}
\author{O. K\'alm\'an${}^1$, Z. Dar\'azs${}^1$, F. Brennecke${}^2$, P. Domokos${}^1$ \\[3pt]
{\footnotesize \it  ${}^1$ Institute for Solid State Physics and Optics, Wigner Research Centre for Physics, Hungarian Academy of Sciences, \\ H-1525 Budapest, P.O. Box 49., Hungary \\
${}^2$ Physikalisches Institut, Universit\"{a}t Bonn, Wegelerstrasse 8,
   53115 Bonn, Germany
}}


\begin{abstract}
Bose-Einstein condensates of ultracold atoms can be used to sense fluctuations of the magnetic field by means of transitions into untrapped hyperfine states. It has been shown recently that counting the outcoupled atoms can yield the power spectrum of the magnetic noise. We calculate the spectral resolution function which characterizes the condensate as a noise measurement device in this scheme.  We use the description of the radio-frequency outcoupling scheme of an atom laser which takes into account the gravitational acceleration. Employing both an intuitive and the exact three-dimensional and fully quantum mechanical approach we derive the position-dependent spectral resolution function for condensates of different size and shape.  
\end{abstract}

\maketitle

\section{Introduction}

Laser cooling and trapping techniques allow nowadays for the preparation of isolated atomic samples \cite{Chu98} at ultracold temperatures (well below 1$\mu$K) where in the case of bosonic atoms a Bose-Einstein condensate (BEC) is formed \cite{Ketterle_Nobel,castin2001}. Employing their interaction with electromagnetic fields, all relevant degrees of freedom of ultracold atoms can be controlled with unprecedented precision \cite{fortagh07,Folman_review,Grimm2000}. For example, the internal atomic state can be manipulated and detected very efficiently by external laser or radio-frequency fields. Therefore, a BEC of trapped atoms can be considered as an ideal probe of weak external fields \cite{Wildermuth2005}.

We have recently shown that counting atoms outcoupled from a magnetically trapped BEC amounts to accessing quantum features in the low-frequency current fluctuations of a nanowire, e.g., a carbon nanotube \cite{Kalman_12}. To this end, the internal hyperfine degree of freedom of  ultracold atoms is interfaced to the current of electrically contacted nanowires with the interaction being mediated by the time-dependent magnetic field. By externally tuning the Zeeman splitting between the hyperfine states, the magnetic field can resonantly induce transitions from the trapped to an untrapped state. These atoms can then be detected, eventually, by single atom resolution \cite{fortagh_detect}.  This measurement scheme is analogous to the radio-frequency (rf) outcoupling scheme of an atomlaser  \cite{Bloch_99,LeCoq_01,Ketterle_97,Ballagh_97,Esslinger_00,Aspect_01,Ottl_05,Robins_13,Schneider_Schenzle_00,Steck_98}.  Later we also showed that,  in conjunction with the magnetic effect of the current on the hyperfine states of the atoms, there is also  a significant back action of the collective atomic hyperfine transitions  on the mechanical oscillation of the current-carrying nanowire \cite{Darazs_14}.  
  
In this paper we revisit the scheme of sensing magnetic field noise by a trapped BEC \cite{Kalman_12}, in the system described in Sec.~\ref{sec:Sys}. As a significant advance to our previous model, in Sec.~\ref{sec:Outcoup} the motion in the gravitational field is taken into account in the derivation of a time- and position-dependent wavefunction of the outcoupled atoms. Gravity is of key importance since the time an outcoupled atom spends in the volume of the condensate is limited by the free-fall and has thus a finite-time broadening effect on the spectral resolution. We first present an intuitive approach in Sec.~\ref{sec:Intu}, and then the complete scattering results in Sec.~\ref{sec:Green}.  In Sec.~\ref{sec:Detect} we show how the magnetic-field noise is related to the number of outcoupled atoms and then we evaluate our results for a nonfactorizable BEC wave function in Sec.~\ref{sec:Numeval}. We conclude in Sec.~\ref{sec:Concl}.

\section{System} \label{sec:Sys}

\begin{figure}[b]
\includegraphics[width=0.8\columnwidth]{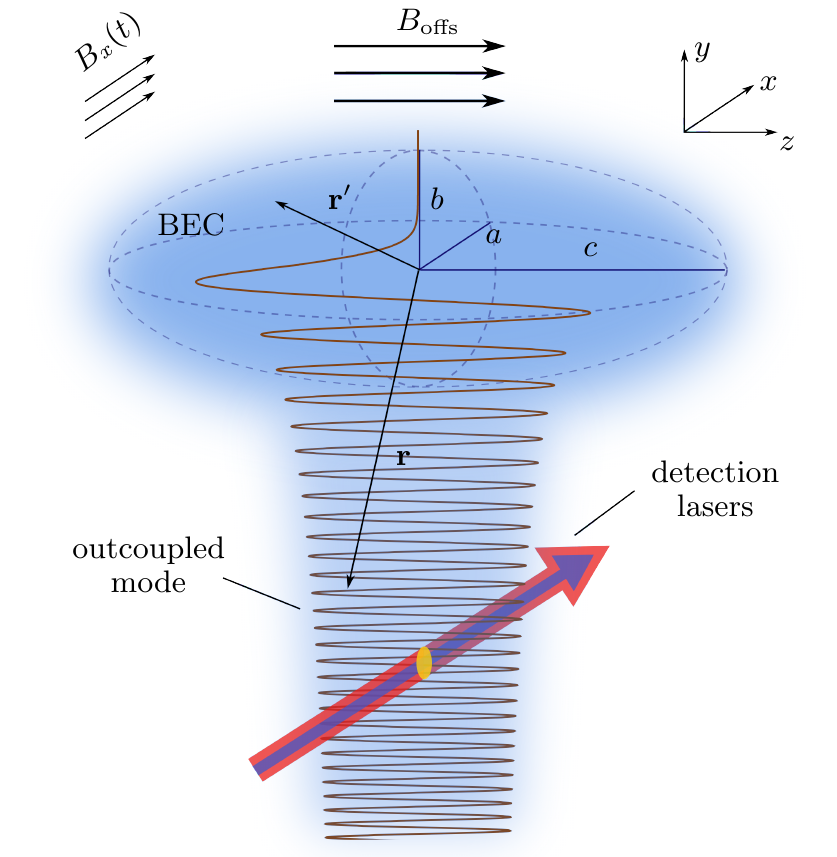}
\caption{(color online) Sketch of the system and the outcoupled mode for a monochromatic outcoupling field.}
\label{Fig1}
\end{figure}

We consider ultracold ${}^{87}$Rb atoms prepared in the ground state hyperfine manifold $F=1$. The total atomic angular momentum $\hat{\mathbf F}$ interacts with the magnetic field according to the Zeeman term $H_{Z} = g_{F} \, \mu_{\mathrm B} \hat{\mathbf F}{\mathbf B}(\rr)$, where $\mu_{\mathrm B} = e\hbar/2 m_{e}$ is the Bohr magneton, the Land\'{e} factor is $g_{F}=-1/2$, and $\hat{\mathbf F}$ is measured in units of $\hbar$. The dominant component of the magnetic field $\mathbf{B(\rr)}$ is considered to be given by a homogeneous offset field $B_{\rm offs}$ pointing along the $z$ direction. The eigenstates of the spin component $\hat F_{z}$, labelled by $m_F=-1,0,1$, are well separated by the Zeeman shift. The inhomogeneous component of the magnetic field $\mathbf{B}(\rr)$ creates a harmonic trapping potential around the minimum of the total magnetic field. In addition, we consider the spin independent static gravitational potential $Mgy$, with atomic mass $M$ and gravitational acceleration $g$ (see Fig.~\ref{Fig1}).

The trap is confining atoms in the low-field seeking state $m_{\mathrm{F}}=-1$ only, and, to a good approximation, gives rise to the static potential $V_{-1}(\rr)=\hbar\omega_{\mathrm{L}}+V_{\mathrm{T}}(\rr)$, where $\omega_{\mathrm{L}}=\frac{1}{\hbar}\left[\left|g_{\mathrm{F}}\right| \mu_{\mathrm{B}}B_{\mathrm{offs}}+\frac{Mg^{2}}{2\omega_{y}^{2}}\right]$ is the  Larmor frequency at the minimum of the potential (chosen as origin) and  $V_{\mathrm T}(\rr)=\frac{M}{2}\left[\omega_{x}^{2}x^{2}+ \omega_{y}^{2}y^{2}+\omega_{z}^{2}z^{2}\right]$ is the harmonic trapping  potential with $\omega_{x}$, $\omega_{y}$ and $\omega_{z}$ being the trap frequencies in the $x$, $y$, and $z$ directions, respectively. Note that the trap center does not coincide with the minimum of the magnetic field but is displaced by a significant gravitational sag $y_{0}=-g/\omega_{y}^{2}$. 

To outcouple atoms from the trapped BEC wavefunction, we consider a spatially homogeneous, time-varying magnetic field  $B_{x}(t)=B\, U(t) \cos(\omega_{\mathrm{rf}}t)$ polarized  in the $x$ direction, with a monochromatic carrier frequency of $\omega_{\sub{rf}}$ in the radio-frequency domain, which is tunable around the Larmor frequency $\omega_{\mathrm{L}}$. We consider this driving field to carry magnetic field noise, 
that is, on top of the carrier frequency $\omega_{\mathrm{rf}}$ there is a time-dependent amplitude,  $U(t)$ being dimensionless. Due to the Zeeman interaction, this magnetic field can quasi-resonantly generate transitions between the magnetic sublevels $m_{\rm F}=-1$ and $m_{\rm F}=0$. 

In the magnetically trapped, $m_{F}=-1$ state,  we assume a pure BEC described by the second quantized field operator $\hat{\Psi}_{-1}(\rr,t)= \sqrt{N} \Phi_{\mathrm{BEC}}(\rr)e^{-i(\omega_{\mathrm{L}}+\mu/\hbar)t}$, where the wavefunction $\Phi_{\mathrm{BEC}}$ is the stationary solution of the Gross-Pitaevskii equation with chemical potential $\mu$ and atom number $N$. Atoms in the Zeeman sublevel $m_{F}=0$, described by the field operator $\hat{\Psi}_{0}(\rr,t)$ are not trapped magnetically and move under the influence of gravity and the mean-field potential $Ng_{s}\Phi_{\mathrm{BEC}}^{2}(\rr)$, with $g_{s}=4\pi\hbar^{2}a_{s}/M$, and scattering length $a_{s}$ ($a_{s}$=5.4 nm for ${}^{87}$Rb). Once spatially separated from the trap, the outcoupled atoms can be detected and counted \cite{Jeltes2007,Bourdel_06,Ottl_05,Ritter_07,Donner_07,HopeClose04}.  We assume that initially no atoms populate the $m_{\rm F}=0$ state. To leading order in  the small quantum field amplitude $\hat{\Psi}_{0}$, the equation of motion  for the $m_{\mathrm{F}}=0$ component in rotating-wave approximation reads
\begin{multline}
\!\!\!i\hbar\frac{\partial}{\partial t}\hat{\Psi}_{0}\!=\! 
\left[-\frac{\hbar^{2}\nabla^{2}}{2M}+Mgy+
N\!g_{s}\!\left|\Phi_{\mathrm{BEC}}(\rr,t)\right|^{2}\right]
\hat{\Psi}_{0} \\ 
-\hbar \, \eta \, \Phi_{\mathrm{BEC}}(\rr) \, U(t) \, e^{i\Delta\cdot t}\,
, \label{time_ev} 
\end{multline}
where $\eta=\mu_{\mathrm{B}} B \sqrt{N}/4\sqrt{2}\hbar$ ($\mu_{\mathrm{B}}$ being the Bohr magneton) and $\Delta=\omega_{\mathrm{rf}}-\omega_{\sub L}-\mu/\hbar$ is the detuning of the radio-frequency from the transition frequency at the trap center. Here we considered the BEC of $m_{\mathrm{F}}=-1$ atoms as an undepleted reservoir from which the $m_{\mathrm{F}}=0$ atoms can be excited by Zeeman transitions. The quantum fluctuation $\delta\hat{\Psi}_{-1}$ is neglected in comparison with $\Phi_{\mathrm{BEC}}$ and the quantum field component in the sublevel $m_{\mathrm{F}}=1$ which is populated via the intermediate $m_F=0$ state is also negligible.  

Within these approximations, the dynamics of the outcoupled field $\hat{\Psi}_{0}(\rr,t)$ decouples from the other Zeeman states. The corresponding partial differential equation in Eq. (\ref{time_ev}), however, cannot be solved analytically. One possible approximation consists in neglecting the kinetic energy term, i.e., neglecting the motion during the outcoupling process. The resulting ordinary differential equation can then be integrated independently for all spatial positions. If the integration time is long enough, the outcoupling process for a  monochromatic excitation takes place from resonant surfaces of constant magnetic field \cite{Steck_98,Gunther_15}, which are close to horizontal planes for a BEC size much smaller than the gravitational sag.

In what follows we will resort to a different approach which accounts for the motion of the outcoupled atoms. It is based on the solution of the quantum mechanical free-fall problem which can be expressed analytically in terms of the Airy functions as eigenfunctions \cite{Harkonen_10,Kramer_06,Schneider_Schenzle_00,Schneider_99}. The additional term in the equation of motion is the mean-field potential $N\!g_{s}\!\left|\Phi_{\mathrm{BEC}}(\rr,t)\right|^{2}$ which varies much less over the size of the condensate than the gravitational potential $Mgy$. While the former varies between $0$ and $\mu$, the latter changes from $-Mgb$ to $Mgb$ along the $y$ direction. With typical experimental parameters for $^{87}$Rb \cite{Gunther_15}, the vertical BEC radius $b$ is on the order of a few $\mu$m resulting in a typical value for $Mgb/\hbar$ of approximately $2\pi\times 6$ kHz, while $\mu/\hbar$ is $2\pi\times 0.5$ kHz. 
Therefore it is much more justified to neglect the mean-field potential than the kinetic energy gained during the outcoupling process. 

\section{Outcoupling} \label{sec:Outcoup}

The outcoupling process for a monochromatic field resonant with the transition frequency at the trap center is sketched in Fig.~\ref{Fig1}. In what follows we will present two approaches for the quantum mechanical description of the problem and determine the outcoupled wave function for a noisy outcoupling field. 

\subsection{Intuitive approach} \label{sec:Intu}

Within the approximation of neglecting the collisions between the outcoupled atoms and the BEC, the most intuitive approach is to expand the solution of Eq. (\ref{time_ev}) in the basis  $\phi_{\left\lbrace k_{x},E_{y},k_{z}\right\rbrace}$ formed by the product of plane waves $\psi_{k_{x}}(x)=e^{ik_{x}x}$ and $\psi_{k_{z}}(z)=e^{ik_{z}z}$ in the horizontal directions, and Airy functions $\mathrm{Ai}$ in the vertical ($y$) direction
\begin{equation}
\psi_{E_{y}}(y)=\frac{1}{l_{0}\!\sqrt{Mg}}\,\mathrm{Ai}\left(\frac{1}{l_{0}}
\left(y-\frac{E_{y}}{Mg}\right)\right),
\label{Airy_basis}
\end{equation}
where $l_{0}=\left(\hbar^{2}/2M^{2}g\right)^{\frac{1}{3}}$ is the 'natural length' of the Airy problem ($l_{0}\approx 0.3$ $\mu$m for $^{87}$Rb). The asymptotics of the functions $\mathrm{Ai}$ satisfy the conditions required for the wave function of a free-falling particle and are orthonormal in the sense that $\braket{\psi_{E_{y}}}{\psi_{E^{\rq}_{y}}}=\int dy \, \psi_{E_{y}}^{\ast}\!(y)\,\psi_{E^{\rq}_{y}}\!(y)=\delta\left(E_{y}-E^{\rq}_{y}\right)$ and form a complete spatial basis $\int dE_{y}\, \psi_{E_{y}}^{\ast}(y)\,\psi_{E_{y}}(y\rq)=\delta(y-y\rq)$ \cite{Harkonen_10}. 

The outcoupled-field operator $\hat{\Psi}_{0}$ can be expanded in terms of the basis functions $\phi_{\left\lbrace k_{x},E_{y},k_{z}\right\rbrace}$ as
\begin{multline}
\hat{\Psi}_{0}(\rr,t)=\frac{1}{(2\pi)^{2}}\!\int\!\! dk_{x}\! \int\!\! dE_{y}\! 
\int \!\! dk_{z} \,  
\hat{c}_{\left\lbrace k_{x},E_{y},k_{z}\right\rbrace}(t) \\
\times 
\phi_{\left\lbrace k_{x},E_{y},k_{z}\right\rbrace}(\rr),
\end{multline}
where $\hat{c}_{\left\lbrace k_{x},E_{y},k_{z}\right\rbrace}(t)$ are the annihilation operators in the respective modes. The time evolution of the amplitudes $\hat{c}_{\left\lbrace k_{x},E_{y},k_{z}\right\rbrace}$ is obtained by integrating the equations of motion \cite{Choi00}
\begin{multline}
\hat{c}_
{\left\lbrace k_{x},E_{y},k_{z}\right\rbrace}(t)=\,
i \,\eta \, \int d^3\rr\rq \phi_{\left\lbrace k_{x},E_{y},k_{z}\right\rbrace}^{\ast}(\rr\rq)\Phi_{\mathrm{BEC}}(\rr\rq) \\
 \times e^{-i\frac{E}{\hbar}t} \int\limits_{0}^{t}\!\!dt\rq \, U(t\rq) \,
 e^{i\,\left(\Delta+\frac{E}{\hbar}\right) t\rq},  
\label{c_eq}
\end{multline}
where $E=\frac{\hbar^{2}}{2M}\left(k_{x}^{2}+k_{z}^{2}\right)+E_{y}$ is the energy pertaining to the basis function $\phi_{\left\lbrace k_{x},E_{y},k_{z}\right\rbrace}$, and we used $\hat{c}_{\left\lbrace k_{x},E_{y},k_{z}\right\rbrace}(0)=0$ (by neglecting zero-point fluctuations). Upon substituting the Fourier decomposition $U(t)=\int_{-\infty}^{\infty}\!d\omega \, \tilde{U}(\omega) \, e^{-i\omega t}$ into Eq. (\ref{c_eq}) and using the identity 
\begin{multline}
\lim_{t\rightarrow\infty}\int\limits_{0}^{t}dt\rq 
e^{i\,\left(\Delta+\frac{E}{\hbar}-\omega\right)\, t\rq}=
2\pi\hbar\, \delta (E-\hbar(\omega-\Delta)) \label{lim_t}
\end{multline}
the time integral can be carried out, and one obtains the frequency-composition of the outcoupled wave
\begin{equation}
\hat{\Psi}_{0}(\rr,t)=i\hbar\eta \!
\int\limits_{-\infty}^{\infty}\!\!d\omega \, \tilde{U}(\omega)
e^{-i(\omega-\Delta)t}
f(\omega-\Delta,\rr), \label{Psi_0}
\end{equation}
where 
\begin{eqnarray}
f(\omega-&&\Delta,\,\rr)\!
=\!\frac{1}{(2\pi)^{2}}\!\!\int\limits_{-\infty}^{\infty}\!\!dk_{x}\!\!
\int\limits_{-\infty}^{\infty}\!\!dE_{y}\!\!
\int\limits_{-\infty}^{\infty}\!\!dk_{z} \,
2\pi\delta\!\left(E\!-\!\hbar(\omega-\Delta)\right)\! 
\notag \\
&&\times 
\int d^3\rr\rq \phi_{\left\lbrace k_{x},E_{y},k_{z}\right\rbrace}^{\ast}(\rr\rq)\Phi_{\mathrm{BEC}}(\rr\rq)
\phi_{\left\lbrace k_{x},E_{y},k_{z}\right\rbrace}(\rr).
\label{f_omega}
\end{eqnarray}

Let us note that the frequency bandwidth associated with the natural length of any resonant Airy function is $Mgl_{0}/\hbar=(mg^2/2\hbar)^{1/3}\approx 2\pi/(1.2 \, \mathrm{ms})$ for $^{87}$Rb. Therefore the integration time of about 10 milliseconds  ensures a good enough frequency resolution, which can be represented by the asymptotic limit taken in Eq. (\ref{lim_t}). We also remark that $\tilde{U}(\omega)$ can represent in principle a fluctuating quantum field, hence the operator character of $\hat{\Psi}_{0}$ is retained. By contrast, the condensate part $\hat{\Psi}_{-1}$ was replaced by the coherent wave function $\Phi_{\mathrm{BEC}}$, thus $f(\omega-\Delta,\rr)$ is simply a c-number. 
 
The numerical evaluation of Eq. (\ref{f_omega}) for a general geometry of the condensate is still challenging. 
To minimize the number of integrals we consider in the following the case of a cylindrically symmetric condensate where the BEC radii $a$ and $c$ perpendicular to the direction of gravity are equal. As shown in Appendix A, $f(\omega-\Delta,\rr)$ can then be written as
\begin{multline} 
\!\!\!f(\omega-\Delta,\rr)=\frac{1}{\sqrt{2}a^{2}}
\!\!\int\limits_{0}^{\infty}\!\!d\bar{k}_{\perp}\bar{k}_{\perp} 
 J_{0}(\bar{k}_{\perp}\bar{r}_{\perp}) \\
\times 
\braket{\phi_{\left\lbrace \bar{k}_{\perp},E_{y}\right\rbrace}\!}{\!\Phi_{\mathrm{BEC}}}
\, \psi_{E_{y}(\bar{k}_{\perp})}(y),
\label{eq:f_omega}
\end{multline}
where $\bar{k}_{\perp}$ is the length of the dimensionless wave vector, and $\bar{r}_{\perp}$ is the length of the dimensionless position vector perpendicular to the direction of gravity,  $E_{y}(\bar{k}_{\perp})=\hbar(\omega\!-\!\Delta)\!-\!\frac{\hbar^{2}}{2M\!a^{2}}\bar{k}_{\perp}^{2}$ is the energy in the $y$ direction, $J_{0}$ is the zeroth order Bessel function, and $\braket{\phi_{\left\lbrace \bar{k}_{\perp},E_{y}\right\rbrace}\!}{\!\Phi_{\mathrm{BEC}}}$ is the scalar product in cylindrical coordinates. This integral represents the linear combination of all the 3D basis functions the energy of which is resonant with a single frequency component $\omega$ of the magnetic field. It is an interesting limit that in 1D only one basis function would be resonant, which leads to the simplified form 
\begin{equation} 
\!\!\!f^{\mathrm{1D}}(\omega-\Delta,y)=
\braket{\psi_{E_{y}}}{\Phi_{\mathrm{BEC}}^{\mathrm{1D}}}
\, \psi_{E_{y}}(y).
\label{f_1D}
\end{equation} 

\subsection{Scattering approach} \label{sec:Green}

In the previous subsection we used a simplified quantum mechanical description of the outcoupling process, which was based on the assumption that the Airy functions $\mathrm{Ai}$ form a complete basis of the problem in the $y$ direction. In fact, this is not accurate. In order to solve the inhomogeneous differential equation (\ref{time_ev}), one has to consider it as a scattering problem and use the Green function of the corresponding free problem to determine the outcoupled wave function.

The solution of Eq. (\ref{time_ev}) is given by
\begin{equation}
\hat{\Psi}_{0}(\rr,t)=i \eta \!\int\limits_{0}^{t} \!\! dt\rq \!\! \int \!\! 
d^{3}\rr\rq 
K(\rr,\rr\rq,t-t\rq)\, U(t\rq) \, e^{i\Delta\cdot t\rq} 
\Phi_{\mathrm{BEC}}(\rr\rq)
\end{equation}  
where $K(\rr,\rr\rq,t-t\rq)$ is the propagator of the free problem (i.e. Eq. (\ref{time_ev}) without the inhomogeneous source term). After substituting the Fourier decomposition of $U(t)$ and taking the limit $t\rightarrow \infty$ one finds that the outcoupled wave function can be expressed in terms of the energy-dependent Green function
\begin{equation}
G^{\mathrm{3D}}(\rr,\rr\rq;\hbar(\omega-\Delta))=\frac{1}{i\hbar}
\int\limits_{0}^{\infty}d\tau K(\rr,\rr\rq,\tau)e^{i(\omega-\Delta)\tau},
\end{equation}
and can be written in the form analogous to Eq. (\ref{Psi_0}) 
\begin{equation}
\hat{\Psi}_{0}(\rr,t)=i\hbar \eta \int\limits_{-\infty}^{\infty}d\omega \, \tilde{U}(\omega) \, e^{-i(\omega-\Delta)t} F(\omega-\Delta,\rr)
\end{equation} 
with  
\begin{equation}
F(\omega-\Delta,\rr)=i\int d^{3}\rr\rq G^{\mathrm{3D}}(\rr,\rr\rq;\hbar(\omega-\Delta))\Phi_{\mathrm{BEC}}(\rr\rq).
\end{equation}
Since the three-dimensional Green function $G^{\mathrm{3D}}$ can be expressed with the Green-function $G^{\mathrm{1D}}$ of the 1D free-fall problem  \cite{Kramer_06} one can write $F(\omega-\Delta,\rr)$ in a form analogous to Eq. (\ref{f_omega}):
\begin{eqnarray}
&&F(\omega\!-\!\Delta,\rr)\!=\!\frac{i}{(2\pi)^{2}}\!\!\!\int\limits_{-\infty}^{\infty}\!\!\!dk_{x}
\!\!\!\int\limits_{-\infty}^{\infty}\!\!\!dk_{z}\, e^{i(k_{x}x+k_{z}z)} \!\!\!
\int \!\! d^{3}\rr\rq \Phi_{\mathrm{BEC}}(\rr\rq) \notag \\
&&\times  e^{-i(k_{x}x\rq+k_{z}z\rq)}\! G^{\mathrm{1D}}\!\left(\!y,y\rq;\hbar(\omega-\Delta)
-\frac{\hbar^{2}}{2M}\left(k_{x}^{2}+k_{z}^{2}\right)\!\!\right) \label{F_Delta_omega}
\end{eqnarray}
where 
\begin{eqnarray}
G^{\mathrm{1D}}(y,y\rq;E)=&&-\frac{\pi}{Mgl_{0}^{2}}
\mathrm{Ai}\left(\frac{y+y\rq+\left|y-y\rq\right|}{2l_{0}}-\frac{E}{Mgl_{0}}\right) \notag  \\
&&\times \mathrm{Ci}\left(\frac{y+y\rq-\left|y-y\rq\right|}{2l_{0}}-\frac{E}{Mgl_{0}}\right),
\label{G_1D}
\end{eqnarray} 
$\mathrm{Ci}$ being the complex Airy function $\mathrm{Ci}(x)=\mathrm{Bi}(x)+i\mathrm{Ai}(x)$ \cite{Kramer_06,Bracher_98,Elberfeld_88} (for a derivation see Appendix \ref{appB}). $G^{\mathrm{1D}}$ fulfills the boundary conditions: For coordinates above the source  ($y>y\rq$) it behaves like the Airy function $\mathrm{Ai}(y)$:
\begin{equation}
G^{\mathrm{1D}}(y\!>\!y\rq;E)=-\frac{\pi}{Mgl_{0}^{2}}
\mathrm{Ai}\!\left(\!\frac{y}{l_{0}}\!-\!\frac{E}{Mgl_{0}}\!\right)\! 
\mathrm{Ci}\!\left(\!\frac{y\rq}{l_{0}}\!-\!\frac{E}{Mgl_{0}}\!\right)
\end{equation}
therefore, it falls off exponentially above the condensate. On the other hand, for coordinates below the source ($y<y\rq$) it behaves like the complex Airy function $\mathrm{Ci}(y)$:
\begin{equation}
G^{\mathrm{1D}}(y\!<\!y\rq;E)=-\frac{\pi}{Mgl_{0}^{2}}
\mathrm{Ai}\!\left(\!\frac{y\rq}{l_{0}}\!-\!\frac{E}{Mgl_{0}}\!\right)\! 
\mathrm{Ci}\!\left(\!\frac{y}{l_{0}}\!-\!\frac{E}{Mgl_{0}}\!\right) \label{G_below}
\end{equation}
which is essentially an outgoing wave for $y\rightarrow -\infty$, since 
\begin{eqnarray}
&&\mathrm{Ai}(y\rightarrow -\infty)\simeq\frac{1}{\sqrt{\pi}}(-y)^{-\frac{1}{4}}\sin\left(\frac{2}{3}(-y)^{\frac{3}{2}}+\frac{\pi}{4}\right), \notag \\
&&\mathrm{Bi}(y\rightarrow -\infty)\simeq\frac{1}{\sqrt{\pi}}(-y)^{-\frac{1}{4}}\cos\left(\frac{2}{3}(-y)^{\frac{3}{2}}+\frac{\pi}{4}\right) . \label{asympt}
\end{eqnarray}

The numerical evaluation of Eq. (\ref{F_Delta_omega}) requires the same effort as that of Eq. (\ref{f_omega}), however, it can also be simplified for the cylindrically symmetric case. Using the dimensionless variables introduced in Appendix \ref{appA}, we find the expression 
\begin{multline}
\!\!\!\!\!F(\omega-\Delta,\rr)\!=\!i l_{0} \int\limits_{0}^{\infty}\!\!\bar{k}_{\perp}d\bar{k}_{\perp} J_{0}\left(\bar{k}_{\perp}\bar{r}_{\perp}\right) \int\limits_{0}^{1} \!\! \bar{r}\rq_{\perp} d\bar{r}\rq_{\perp} J_{0}\left( \bar{k}_{\perp} \bar{r}\rq_{\perp} \right) \\
\!\!\!\!\times \!\!\!\!\! \int\limits_{-\bar{b}\sqrt{1-\bar{r}\rq^{2}_{\perp}}}^{\bar{b}\sqrt{1-\bar{r}\rq^{2}_{\perp}}} \!\!\!\! d\bar{y}\rq \Phi_{\mathrm{BEC}}(\bar{r}\rq_{\perp},\bar{y}\rq) G^{\mathrm{1D}}\!\left(\!\bar{y},\bar{y}\rq;E_{y}(\bar{k}_{\perp}\!)\!\right) \!\!\!\!\!\!\! \label{eq:F_omega}
\end{multline}
which is analogous to Eq. (\ref{eq:f_omega}).

\section{Detection} \label{sec:Detect}

The density of outcoupled atoms at a position $\rr$ is
\begin{eqnarray}
N\!&& \left(\Delta,\rr\right)=\left< \hat{\Psi}_{0}^{\dagger}(\rr,t)\,
\hat{\Psi}_{0}(\rr,t)\right>=(\hbar \eta)^{2} \!\!\!\int\limits_{-\infty}^{\infty}\!\!\!d\omega f^{\ast}(\omega-\Delta,\rr)
\nonumber \\
&&\times \!\!\!\int\limits_{-\infty}^{\infty}\!\!\!d\omega\rq e^{i(\omega-\omega\rq)t} 
f(\omega\rq-\Delta,\rr)
\left<\!\tilde{U}^{\ast}(\omega)
\tilde{U}(\omega\rq)\!\right> .
\end{eqnarray}
This equation holds also for the approach presented in Section \ref{sec:Green} if one replaces $f(\omega-\Delta,\rr)$ by $F(\omega-\Delta,\rr)$. We note that $N(\Delta,\rr)$ is normalized in the asymptotic limit $t\rightarrow \infty$ as a rate of atoms per unit time.

Let us assume that the magnetic-field noise is incoherent, i.e., $\left<\tilde{U}^{\,\ast}(\omega)\, \tilde{U}(\omega\rq)\right>=S(\omega\rq)\delta(\omega-\omega\rq)$, where $S(\omega)=\int_{-\infty}^{\infty}d\tau e^{i\omega\tau}\left<U(0)U(\tau)\right>$ is the magnetic noise spectrum (or power spectrum of the magnetic field). Then the density of outcoupled atoms per unit time can be written as 
\begin{equation}
N\!\left(\Delta,\rr\right)= \left(\frac{\hbar\eta}{Mgl_{0}}\right)^{\!2}
\int\limits_{-\infty}^{\infty}\!d\omega\,D(\omega-\Delta,\rr)\,S(\omega)
\label{N_Delta}
\end{equation}
where $D(\omega-\Delta,\rr)=(Mgl_{0})^{2}\left|f(\omega-\Delta,\rr)\right|^{2}$ is the \emph{spectral resolution function} of the BEC employed as a measuring device. We note that other sources of magnetic field noise can also be present in  experiments, for instance that of the offset magnetic field \cite{Fauquembergue_05}, which can also be described with a convolution formula very similar to Eq. (\ref{N_Delta}). $D(\omega-\Delta,\rr)$ is a density function in coordinate space with the dimension of $1/\mathrm{Volume}$, and it depends parametrically on the frequency $\omega-\Delta$. This parametric dependence can be determined experimentally \cite{Gunther_15} using a tunable monochromatic outcoupling field for which $S(\omega)\sim \delta(\omega)$: The number of outcoupled atoms as the function of the frequency of the outcoupling field readily gives $D$. In the following, we will evaluate the spectral resolution function for different trapping geometries.

\section{Numerical evaluation} \label{sec:Numeval}

In order to simplify the further calculations, we will use the Thomas-Fermi solution for the BEC wavefunction,  
\begin{equation}
\Phi_{\mathrm{BEC}}(\rr)=\sqrt{\left[\mu-V_{\mathrm{T}}(\rr)\right]/Ng_{s}}\,,
\label{eq:Phi_BEC}
\end{equation}
where $\mu=\left(15Ng_{s}\omega_{x}^{2}\omega_{y}/8\pi\right)^{2/5}
\left(M/2\right)^{3/5}$ is the chemical potential.
The condensate has thus an ellipsoidal shape with a parabolic density distribution, and the form of Eqs. (\ref{eq:f_omega}) and (\ref{eq:F_omega}) is derived in Appendix \ref{appC}. 

\begin{figure}[tbh]
\subfigure{
\includegraphics[width=1.0\columnwidth]{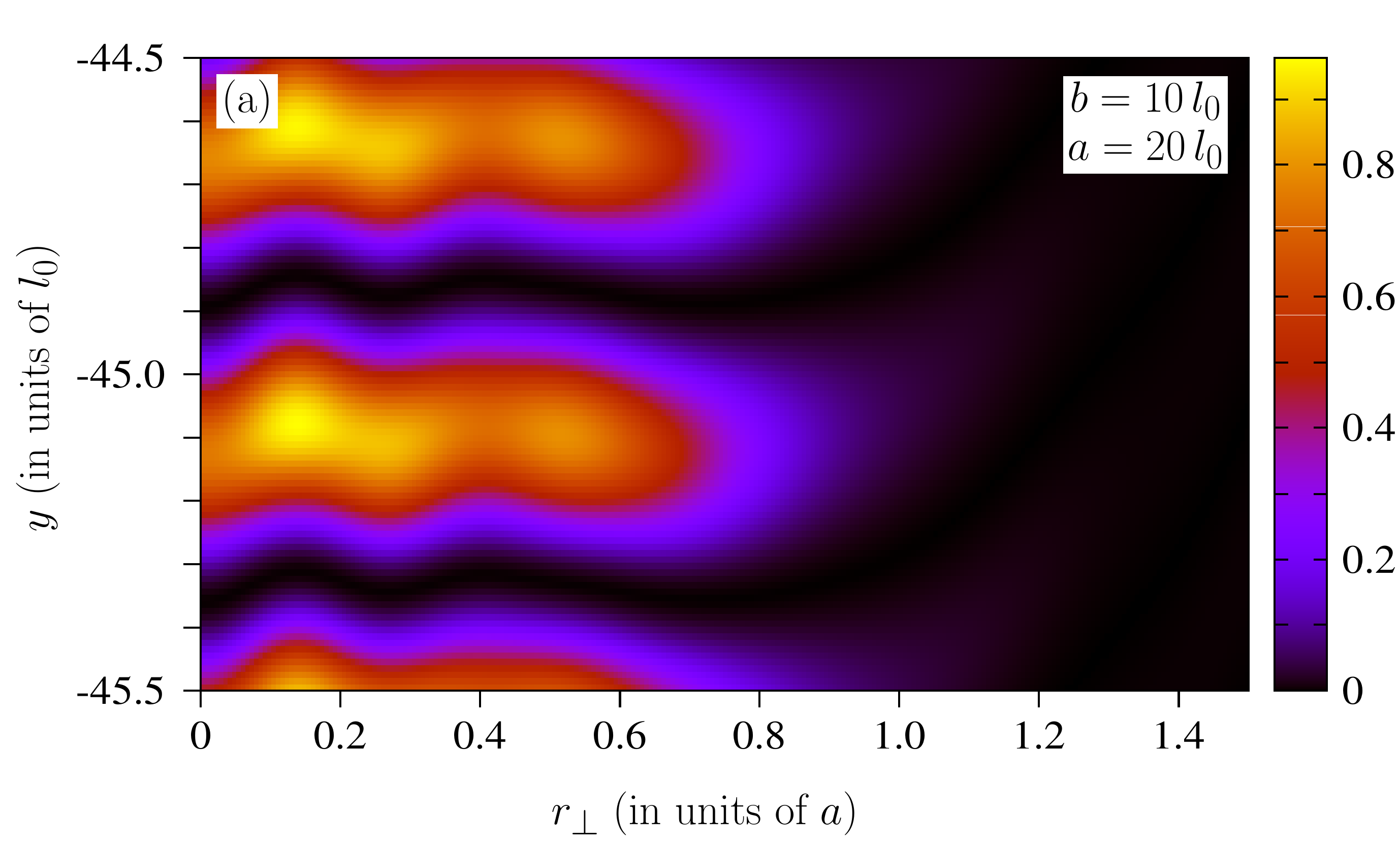}
\label{Fig2a}} \\
\vspace{-3.5ex}
\subfigure{
\includegraphics[width=1.0\columnwidth]{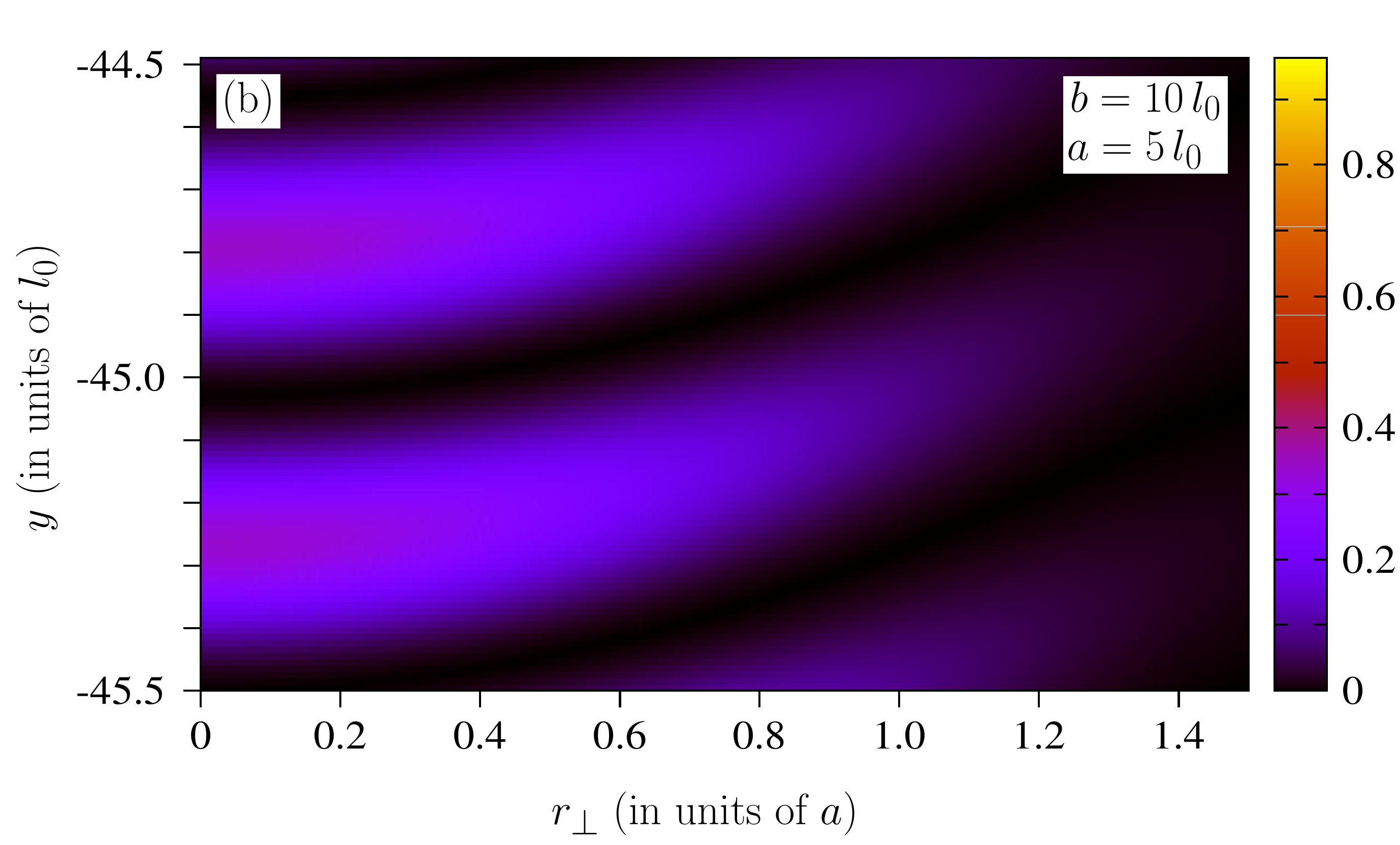}
\label{Fig2b}} \\
\vspace{-3.5ex}
\subfigure{
\includegraphics[width=0.9\columnwidth]{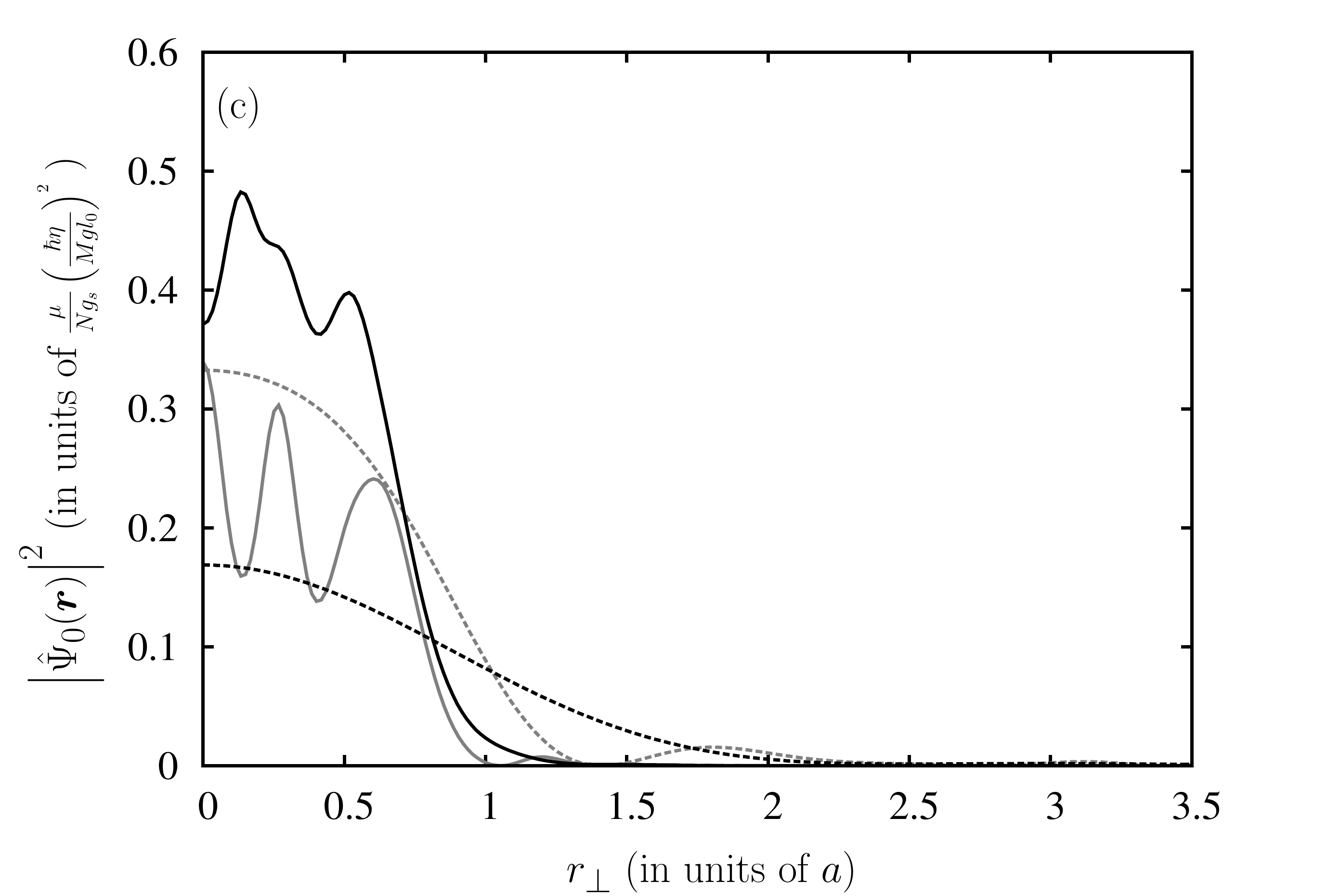}
\label{Fig2c}}
\caption{(color online) The outcoupled atom beam profile by the results of Sec. \ref{sec:Intu} representing both the vertical and the radial directions for a BEC with semi-axes $a=20\,l_{0}$ and $b=10\,l_{0}$ (a), and $a=5\,l_{0}$ and $b=10\,l_{0}$ (b) around the spatial coordinate $y=-45\,l_{0}$, for $\omega-\Delta=0$. (c) The radial distribution of atomic density
at $y=-45.25\,l_{0}$. The grey and black curves correspond to the 'intuitive' and 'complete' approaches presented in Secs. \ref{sec:Intu}, and \ref{sec:Green}, the solid and dashed curves to cases (a) and (b), respectively.}
\label{Fig2}
\end{figure}

Let us note that the above assumption for the BEC wavefunction does not allow for a factorization of the outcoupled wavefunction as a product of terms varying in the orthogonal spatial directions. The vertical and horizontal dynamics are intricately coupled, which is  exhibited by the two-dimensional plots in Fig.~\ref{Fig2a} and \ref{Fig2b} showing the atomic density $\langle \hat{\Psi}_0^\dag \hat{\Psi}_0\rangle$ at a given position below the source. Although the scattering approach of Sec. \ref{sec:Green} provides the complete description, it is instructive to show the results obtained from the intuitive approach of Sec. \ref{sec:Intu}. This illustrates how a limited set is selected from the basis of Airy functions by a monochromatic rf driving field. The fast oscillations, which are characteristic of the Airy function $\mathrm{Ai}(y)$ occur along the vertical direction $y$ (two periods are plotted) because only a narrow band of Airy functions is excited. 

\begin{figure}[tbh]
\includegraphics[width=0.9\columnwidth]{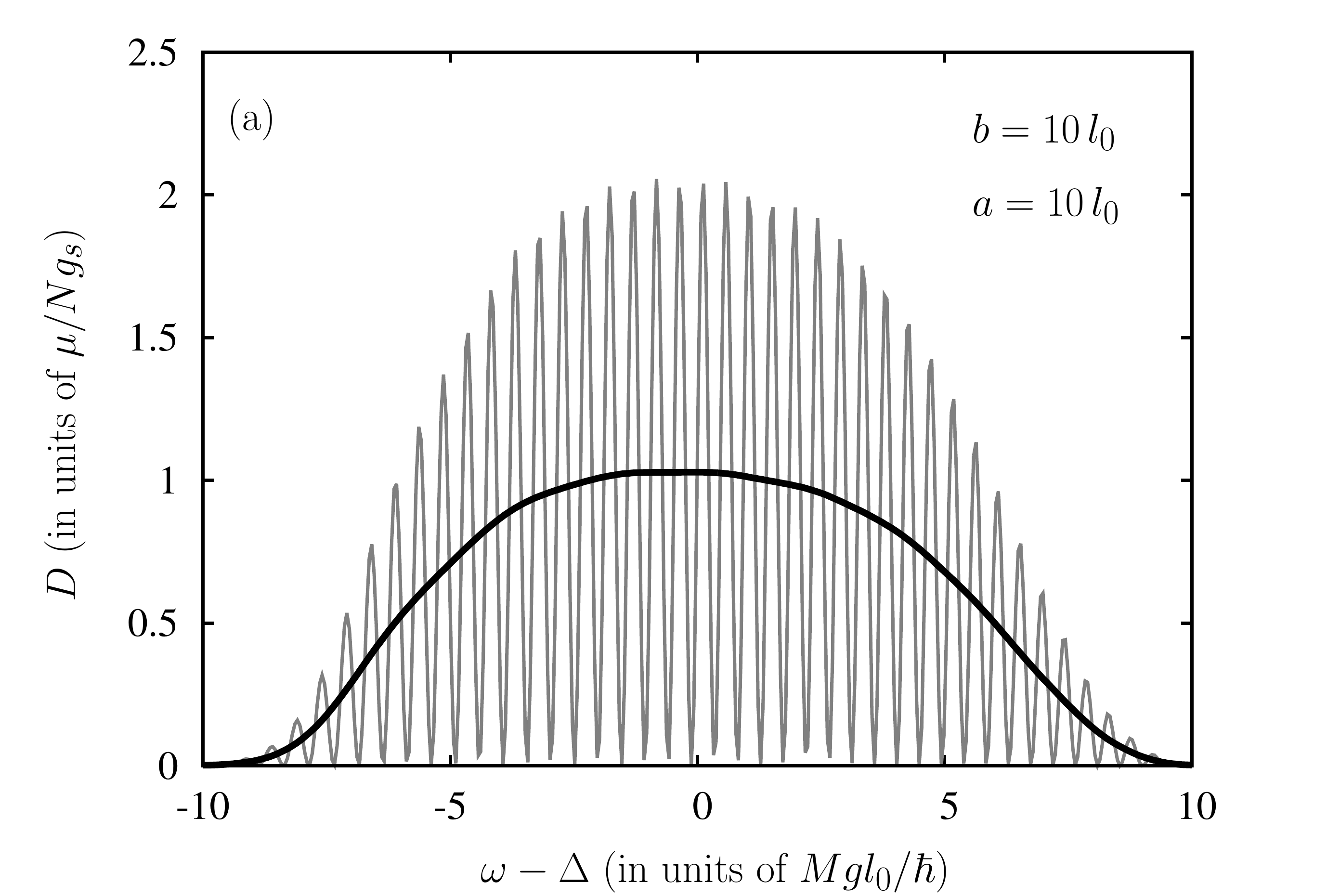} \\
\includegraphics[width=0.9\columnwidth]{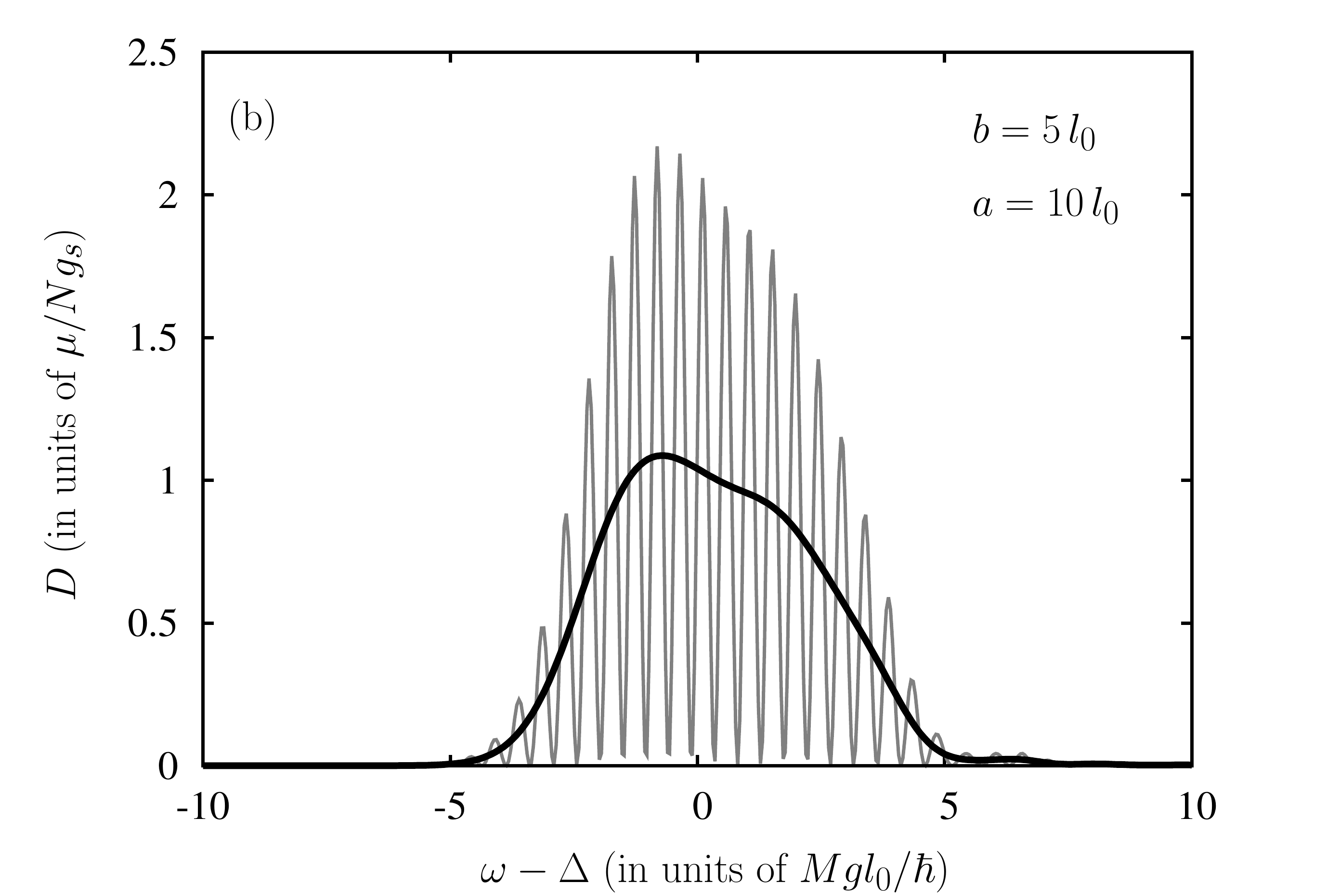} \\
\includegraphics[width=0.9\columnwidth]{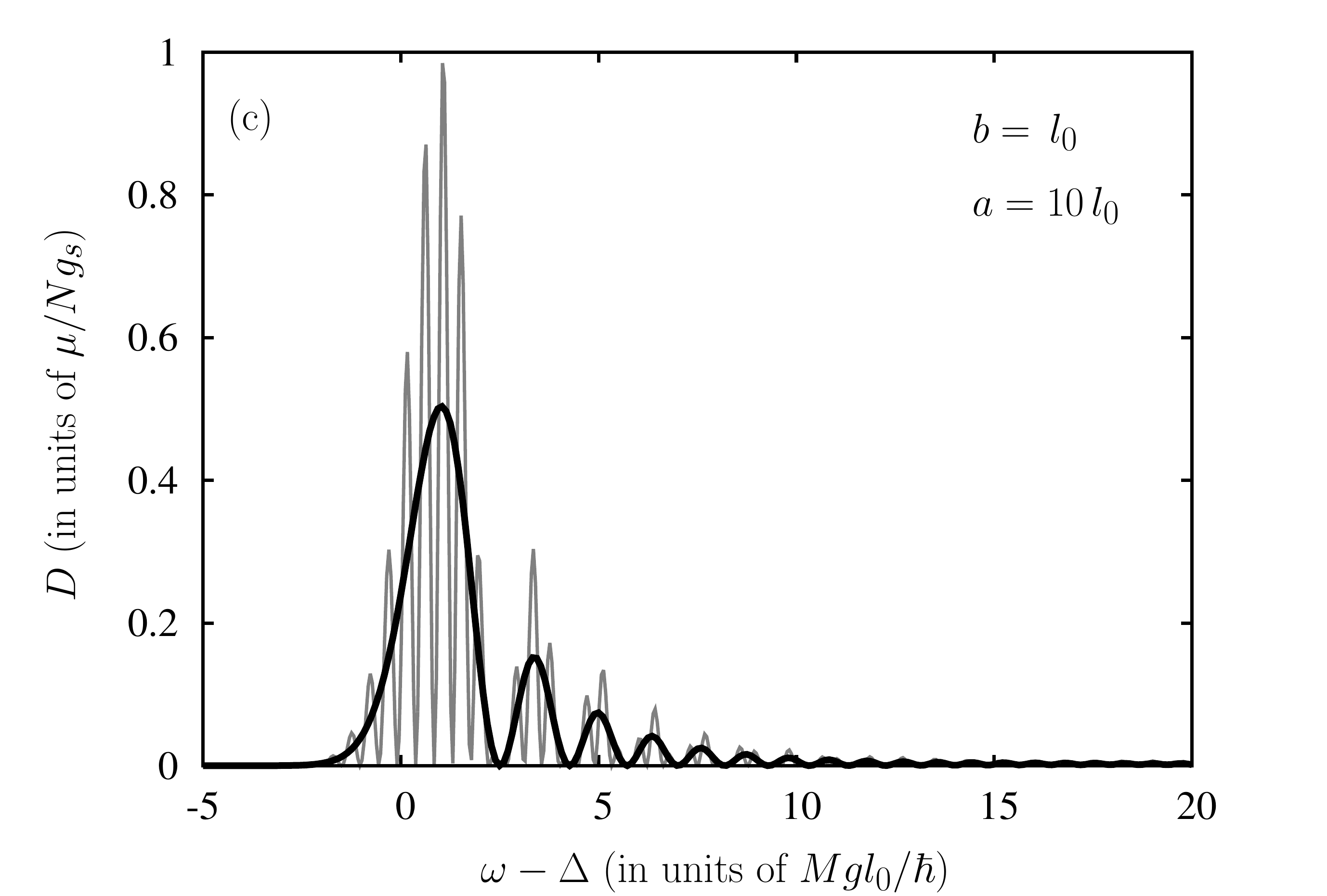}
\caption{The spectral resolution function $D(\omega-\Delta,\rr)$ 
of the BEC as a function of the frequency $\omega-\Delta$  at the spatial coordinate $y=-45\,l_{0}$ on the symmetry axis ($x=0$ and $z=0$) for a cloud with a decreasing vertical semi-axis $b$. The grey and black curves correspond to the 'intuitive' and 'complete' approaches.}
\label{Fig3}
\end{figure}

The radial structure along the coordinate $r_{\perp}$ depends strongly on the condensate shape, which can be seen in both approaches. First, the narrower the source along the transverse directions, the more significant the radial diffraction, c.f. the beam originating from the cigar-shaped condensate. Second, a condensate with larger horizontal extension gives rise to radial excitations including more transverse modes and thus more structures in the radial distribution. 

According to Eq.~(\ref{N_Delta}), measuring the number of atoms as a function of $\Delta$ by varying the Larmor frequency with the offset magnetic field, allows to determine the magnetic noise spectrum with a resolution given by the width of the function $D(\omega-\Delta,\rr)$. The width of $D$ is primarily determined by the size of the condensate in the direction of gravity, as can be seen in Fig.~\ref{Fig3}, which compares $D$ for three different values of the vertical semi-axis $b$ while keeping the horizontal extension constant. The grey and black curves correspond to the  approaches of Sections \ref{sec:Intu}, and \ref{sec:Green}, i.e., Eqs. (\ref{eq:f_omega}) and (\ref{eq:F_omega}), respectively. Going from top to the bottom, the condensate transforms from a spherical to a compressed pancake shape with aspect ratio 1:10. The rapid oscillations of the grey curves reflect the form of the Airy function $\mathrm{Ai}(y)$ which represent the gravitational acceleration until the detection point $\rr$. These oscillations are not present in the case of the black curves, where the function $\mathrm{Ci}(y)$ is of relevance. Here the oscillations of the two types of Airy functions $\mathrm{Ai}$ and $\mathrm{Bi}$ cancel each other, like sine and cosine waves in accordance with the asymptotic forms Eq.~(\ref{asympt}). Note however that the results of the intuitive approach agree with the complete scattering approach, when averaging spatially over a finite detection volume. (Because of the perfect overlap with the thick black curves we do not show the result of such averaging.) 

\begin{figure}[tbh]
\includegraphics[width=0.9\columnwidth]{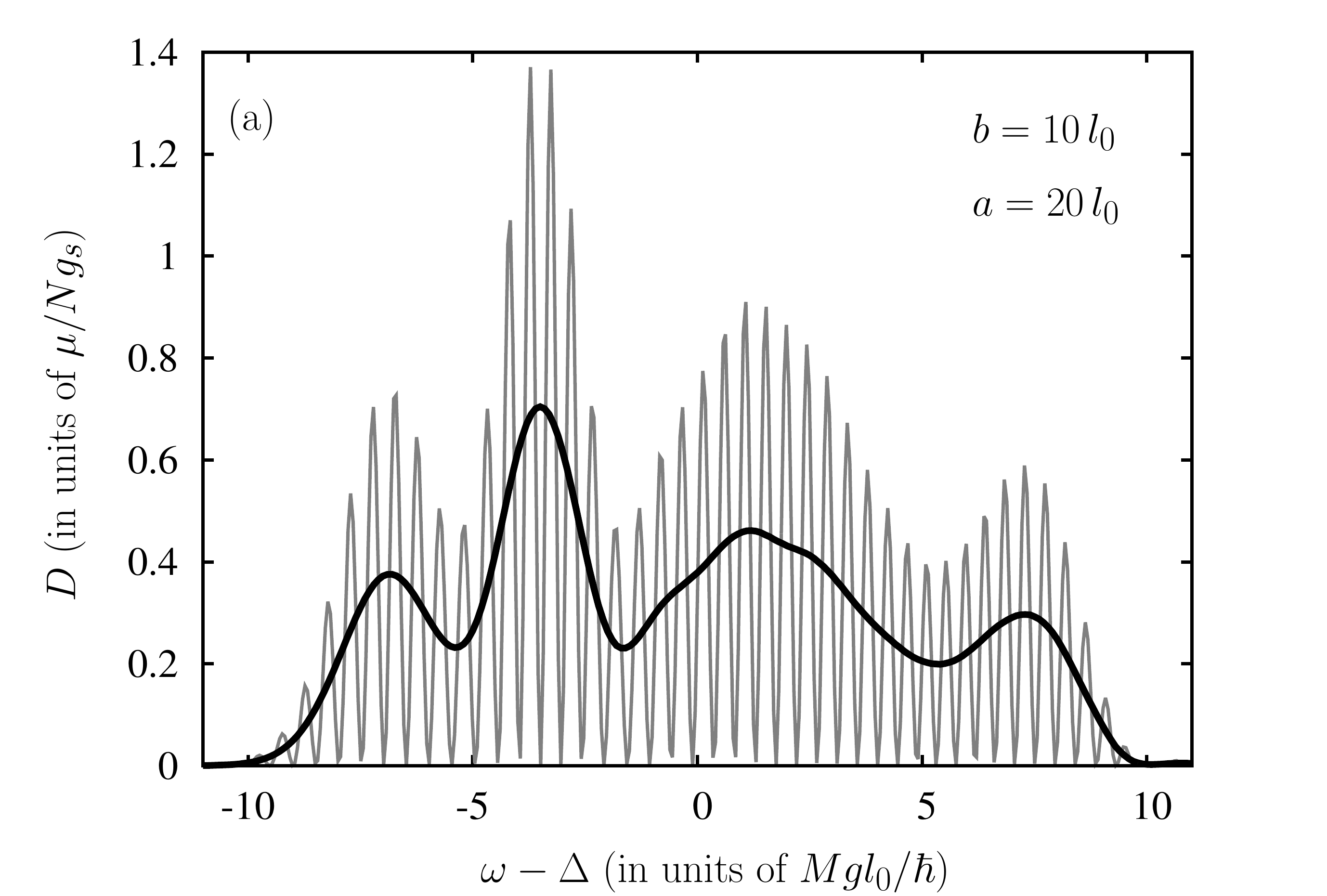} \\
\includegraphics[width=0.9\columnwidth]{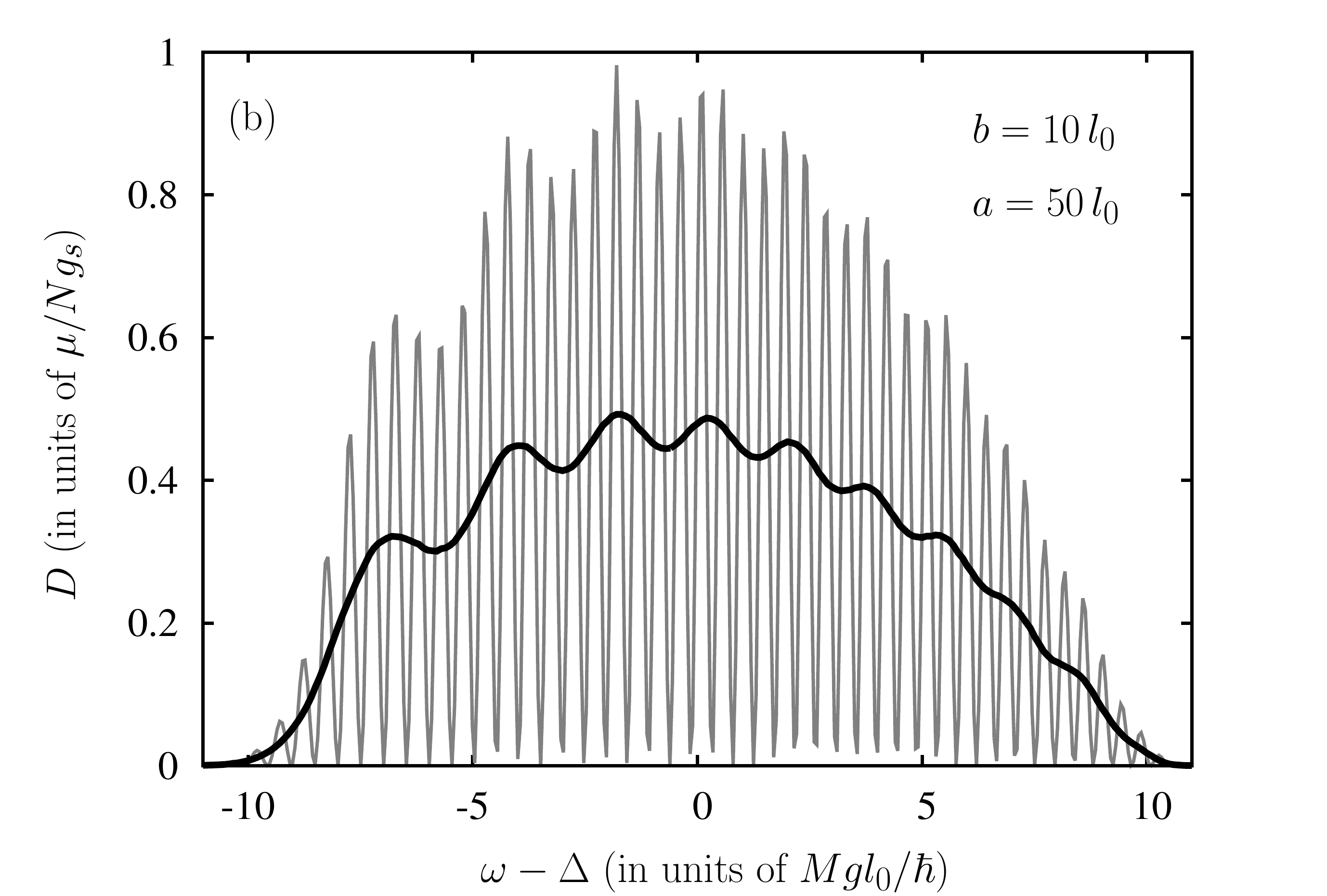} 
\caption{The spectral resolution function $D(\omega-\Delta,\rr)$ 
of the BEC as a function of the frequency $\omega-\Delta$ at the spatial 
coordinate $y=-45\,l_{0}$ on the symmetry axis ($x=0$ and $z=0$) for a cloud with an increasing horizontal radius $a$.  The grey and black curves correspond to the 'intuitive' and 'complete' approaches.}
\label{Fig4}
\end{figure}

The significant deviation of the spectral resolution function in Fig.~\ref{Fig3} from a symmetrical form is connected to the fact that for thin condensates, i.e.~when $b$ becomes comparable to the characteristic length scale $l_{0}$ of the Airy function, it is the Airy function $\mathrm{Ai}\left(\bar{y}\rq\right)$ inside the spatial integral which dominates in Eq. (\ref{eq:f_omega}) and also in Eq. (\ref{eq:F_omega}), therefore, the spectral resolution function inherits the oscillations of the Airy function $\mathrm{Ai}$.

In the case of condensates that are much larger than $l_{0}$ in the vertical direction, different horizontal sizes lead to different structures in the spectral resolution functions, as can be seen in Fig.~\ref{Fig4}. This slowly varying structure arises as a consequence of the decreased energy of the radially excited modes in the enlarged condensate. The broader the source of the atom laser beam, the more radial excitations can be involved to complement the vertical potential energy to fulfill the resonance condition. The multi-mode dynamics in the horizontal direction can also be seen in Fig.~\ref{Fig2a} and \ref{Fig2c}.

\section{Conclusion} \label{sec:Concl}

In conclusion, we have determined the spectral resolution function of magnetically trapped Bose-Einstein condensates, which characterizes its detection properties as a measuring device for magnetic field noise. To provide insight into the outcoupling mechanism, we used an intuitive approach which takes into account the motion of the atoms in the gravitational field. We also presented the complete, three-dimensional solution using the Green function which takes into account the proper boundary conditions as well. The quantum mechanical solution for the outcoupled matter wave allows for constructing the local, position-dependent spectral resolution function in both approaches, which can then be used to describe the detection process for arbitrary spatial resolution.  Beyond the mainly linear dependence on the vertical size (the size in the direction of the gravitational field), the spectral resolution function exhibits a remarkable dependence also on the lateral extension of the condensate, which results from an interplay between the excitation of the different radial modes and the non-factorizability of the BEC wave function.
 
\begin{acknowledgments}
We appreciate helpful discussions with S. Varr\'{o} and T. Kiss. This work was supported by the Hungarian Academy of Sciences (Lend\"ulet Program, LP2011-016) and the National Research, Development and Innovation Office (K115624). O. K. acknowledges support from the J\'{a}nos Bolyai Research Scholarship of the Hungarian Academy of Sciences. F.B. acknowledges support from the DFG.
\end{acknowledgments}

\appendix
\section{Cylindrically symmetric condensate} \label{appA}

In the case of a cylindrically symmetric condensate, with equal semi-axes perpendicular to the direction of gravity ($a\!=\!c\,$), one can introduce a cylindrical coordinate system and define the dimensionless perpendicular wave vector 
$\bar{\boldsymbol{k}}_{\perp}=(k_{x}a,0,k_{z}a)$, with 
$\bar{k}_{\perp}^{2}=a^{2}(k_{x}^{2}+k_{z}^{2})=2Ma^{2}(E_{x}+E_{z})/\hbar^{2}$. In this way the integrals for $k_{x}$ and $k_{z}$ can be substituted by $(1/a^{2})\int_{0}^{\infty}d\bar{k}_{\perp}\bar{k}_{\perp}\int_{0}^{2\pi}d\beta$, where $\beta$ is the angle between the perpendicular wave vector and the horizontal coordinate axis. At the same time, one can also introduce a perpendicular dimensionless position vector $\bar{\rr}_{\perp}=(x/a,0,z/a)$, then Eq. (\ref{f_omega}) becomes 
\begin{eqnarray}
f(\omega-\Delta,\rr)=&&
\frac{1}{(2\pi)^{2}a^{2}}\!\!\int\limits_{0}^{\infty}\!\!d\bar{k}_{\perp}\bar{k}_{\perp}\!
\int\limits_{0}^{2\pi}\!\!d\beta 
e^{i\bar{k}_{\!\perp}
\bar{r}_{\!\perp}\!\cos(\beta-\gamma)}
\nonumber \\ &&\times 2\pi\!\!\!\int\limits_{-\infty}^{\infty}\!\!dE_{y}\,
\delta\!\left(E_{\perp}+E_{y}\!-\!\hbar(\omega-\Delta)\right)\! \nonumber \\
&&\times
\braket{\phi_{\left\lbrace \bar{k}_{\perp},E_{y}\right\rbrace}\!}{\!\Phi_{\mathrm{BEC}}}\, \psi_{E_{y}}(y) \label{f_omega_Delta}
\end{eqnarray}  
where $\gamma$ is the angle between the perpendicular position vector $\bar{\rr}_{\perp}$ and the horizontal coordinate axis, and $E_{\perp}=E_{x}+E_{z}$ is the energy in the perpendicular direction. As the condensate is cylindrically symmetric, the scalar product $\braket{\phi_{\left\lbrace \bar{k}_{\perp}, E_{y}\right\rbrace}\!}{\!\Phi_{\mathrm{BEC}}}$ is independent of the angle $\beta$, and the integral for $\beta$ can be evaluated to be $2\pi\,J_{0}(\bar{k}_{\perp}\bar{r}_{\perp})$, for any $\gamma$.

Because of the Dirac delta, the energy in the $y$ direction has to be equal to $E_{y}=\hbar(\omega-\Delta)-E_{\perp}=\hbar(\omega-\Delta)-\hbar^{2}\bar{k}_{\perp}^{2}/2Ma^{2}$, leading to Eq. (\ref{eq:f_omega}).

\section{Derivation of the one-dimensional Green function} \label{appB}

In order to derive the form of the one-dimensional Green function given in Eq. (\ref{G_1D}) one starts from the 1D free-fall problem:
\begin{equation}
\left(E+\frac{\hbar^{2}}{2M}\frac{\partial^{2}}{\partial y^{2}}-Mgy\right)G^{\mathrm{1D}}(y,y\rq;E)=\delta(y-y\rq).
\end{equation} 
Introducing the dimensionless coordinates 
\begin{equation}
\xi=\frac{1}{l_{0}}\left(y-\frac{E}{Mg}\right), \qquad
\xi\rq=\frac{1}{l_{0}}\left(y\rq-\frac{E}{Mg}\right),
\end{equation}
leads to
\begin{equation}
\left(\frac{\partial^{2}}{\partial \xi^{2}}-\xi\right)G^{\mathrm{1D}}\left(\xi,\xi\rq;E\right)=\frac{2Ml_{0}}{\hbar^{2}}\delta\left(\xi-\xi\rq \right).
\end{equation}
The linearly independent solutions of this equation are the Airy functions $\mathrm{Ai}$ and $\mathrm{Bi}$. In order to satisfy the boundary conditions, the solution has to i) decay exponentially for field coordinates $y$ larger than the source coordinate $y\rq$, and ii) has to behave like an outgoing wave for field coordinates $y$ smaller than the source coordinate $y\rq$. Therefore, we may look for the solution in the form \cite{Bracher_98,Elberfeld_88}:
\begin{equation}
G^{\mathrm{1D}}\left(\xi,\xi\rq;E\right)=a_{>}\Theta\left(\xi-\xi\rq\right)\mathrm{Ai}(\xi)+a_{<}\Theta\left(\xi\rq-\xi\right)\mathrm{Ci}(\xi), \, \label{G_1D_probe}
\end{equation}
where $\Theta(x)$ is the Heaviside function and $\mathrm{Ci}$ is the complex Airy function $\mathrm{Ci}(x)=\mathrm{Bi}(x)+i\mathrm{Ai}(x)$. The coefficients $a_{>}$ and $a_{<}$ can be determined by requiring $G^{\mathrm{1D}}$ and its derivative to be continuous at the point $\xi=\xi\rq$ \cite{Cavalcanti_03}:
\begin{eqnarray}
&&\lim_{\varepsilon\rightarrow 0}G^{\mathrm{1D}}(\xi\rq+\varepsilon,\xi\rq;E) = \lim_{\varepsilon\rightarrow 0}G^{\mathrm{1D}}(\xi\rq-\varepsilon,\xi\rq;E), \nonumber \\
&&\lim_{\varepsilon\rightarrow 0}\left.\frac{\partial G^{\mathrm{1D}}}{\partial\xi}\right|_{\xi=\xi\rq+\varepsilon}- \lim_{\varepsilon\rightarrow 0}\left.\frac{\partial G^{\mathrm{1D}}}{\partial\xi}\right|_{\xi=\xi\rq-\varepsilon}=\frac{2Ml_{0}}{\hbar^{2}}, \,
\end{eqnarray}
leading to the following equations:
\begin{eqnarray}
a_{>} \mathrm{Ai}\left(\xi\rq\right)&&-a_{<} \mathrm{Ci}\left(\xi\rq\right)=0, \nonumber \\
a_{>} \mathrm{Ai}\rq\left(\xi\rq\right)&&-a_{<} \mathrm{Ci}\rq\left(\xi\rq\right)=\frac{2Ml_{0}}{\hbar^{2}}.
\end{eqnarray}
Using the identity $\mathrm{Ai}(x)\mathrm{Bi}\rq(x)-\mathrm{Ai}\rq(x)\mathrm{Bi}(x)=1/\pi$ one can easily determine $a_{>}$ and $a_{<}$. By introducing $u=\xi+\xi\rq$ and $v=\xi-\xi\rq$, the Green function of Eq. (\ref{G_1D_probe}) takes the form
\begin{equation}
G^{\mathrm{1D}}(u,v;E)=-\frac{2Ml_{0}}{\hbar^{2}}\pi \mathrm{Ai}\left( \frac{u+\left|v\right|}{2} \right) \mathrm{Ci} \left( \frac{u-\left|v\right|}{2}\right),
\end{equation}
from which one can readily retrieve the form presented in Eq. (\ref{G_1D}).

\section{The formulas for a cylindrically symmetric condensate} \label{appC}
 
Since $\Phi_{\sub{BEC}}$ of Eq. (\ref{eq:Phi_BEC}) has a finite support (it is nonzero only in the ellipsoid defined by the semi-axes $a$ and $b$), and the condensate is cylindrically symmetric, the scalar product of Eq. (\ref{eq:f_omega}) can be written as
\begin{widetext}
\begin{equation}
\braket{\phi_{\left\lbrace \bar{k}_{\perp},E_{y}\right\rbrace}\!}{\!\Phi_{\mathrm{BEC}}}\!=\! 
2\pi a^{2} l_{0}\sqrt{\!\frac{\mu}{N\!g_{s}}}\!
\int\limits_{0}^{1}\!\!d\bar{r}\rq_{\perp}\bar{r}\rq_{\perp} 
J_{0}(\bar{k}_{\perp}\bar{r}\rq_{\perp})
\int\limits_{-\bar{b}\sqrt{1-\bar{r}\rq^{2}_{\perp}}}^{\bar{b}\sqrt{1-\bar{r}\rq^{2}_{\perp}}}
d\bar{y}\rq 
\psi_{E_{y}(\bar{k}_{\perp})}(\bar{y}\rq)
\sqrt{1-\bar{r}\rq^{2}_{\perp}-\frac{\bar{y}\rq^{2}}{\bar{b}^{2}}},
\end{equation}
\end{widetext}
where $J_{0}$ is the zeroth order Bessel function. Therefore Eq. (\ref{f_omega})
can be written as
\begin{widetext}
\begin{multline}
\!\!\!f(\omega-\Delta,\rr)=\frac{2\pi}{\sqrt{2}}\frac{1}{Mgl_{0}} \sqrt{\!\frac{\mu}{N\!g_{s}}}
\int\limits_{0}^{\infty}\!\!d\bar{k}_{\perp}\bar{k}_{\perp} 
 J_{0}(\bar{k}_{\perp}\bar{r}_{\perp})
\, \mathrm{Ai}\!\left(\bar{y}-\frac{E_{y}(\bar{k}_{\perp})}{Mgl_{0}}\right)
\int\limits_{0}^{1}\!\!d\bar{r}\rq_{\perp}\bar{r}\rq_{\perp} 
J_{0}(\bar{k}_{\perp}\bar{r}\rq_{\perp}) \\
\times
\int\limits_{-\bar{b}\sqrt{1-\bar{r}\rq^{2}_{\perp}}}^{\bar{b}\sqrt{1-\bar{r}\rq^{2}_{\perp}}}
d\bar{y}\rq 
\sqrt{1-\bar{r}\rq^{2}_{\perp}-\frac{\bar{y}\rq^{2}}{\bar{b}^{2}}}
\mathrm{Ai}\!\left(\bar{y}\rq-\frac{E_{y}(\bar{k}_{\perp})}{Mgl_{0}}\right).
\end{multline}
\end{widetext}
Using the same variables and the form (\ref{G_below}) of the 1D Green function valid for coordinates below the condensate, Eq. (\ref{eq:F_omega}) gives 
\begin{widetext}
\begin{multline}
\!\!\!F(\omega-\Delta,\rr)=-\frac{2\pi}{2}\frac{1}{Mgl_{0}} \sqrt{\!\frac{\mu}{N\!g_{s}}}
\int\limits_{0}^{\infty}\!\!d\bar{k}_{\perp}\bar{k}_{\perp} 
 J_{0}(\bar{k}_{\perp}\bar{r}_{\perp}) \, \mathrm{Ci}\!\left(\bar{y}-\frac{E_{y}(\bar{k}_{\perp})}{Mgl_{0}}\right)
\int\limits_{0}^{1}\!\!d\bar{r}\rq_{\perp}\bar{r}\rq_{\perp} 
J_{0}(\bar{k}_{\perp}\bar{r}\rq_{\perp}) \\
\times
\int\limits_{-\bar{b}\sqrt{1-\bar{r}\rq^{2}_{\perp}}}^{\bar{b}\sqrt{1-\bar{r}\rq^{2}_{\perp}}}
d\bar{y}\rq 
\sqrt{1-\bar{r}\rq^{2}_{\perp}-\frac{\bar{y}\rq^{2}}{\bar{b}^{2}}}
\mathrm{Ai}\!\left(\bar{y}\rq-\frac{E_{y}(\bar{k}_{\perp})}{Mgl_{0}}\right).
\end{multline}
\end{widetext}

\bibliographystyle{unsrtnat}

\bibliography{atomlaser_bib} 

\end{document}